\documentclass[12pt]{article}
\usepackage{latexsym}
\usepackage{amsmath}
\usepackage{amsfonts}
\usepackage{amssymb}
\usepackage{graphicx}
\usepackage{epstopdf}

\usepackage{color}
\usepackage{tikz}
\usepgflibrary{arrows}

\title{h-BN layer induced chiral decomposition in the electronic properties of multilayer graphene}
\author{I. Zasada$^1$\thanks {e-mail:izasada@wfis.uni.lodz.pl}, A. Molenda$^1$, P. Ma\'slanka$^2$, K. \L uczak $^1$\\
$^1$\small Department of Solid State Physics,
 Faculty of Physics and Applied Informatics\\
$^2$\small Department of  Computer Science, 
 Faculty of Physics and Applied Informatics\\
\small University of \L\'od\'z,\\
\small Pomorska 149/153, 90-236 {\L}\'od\'z, Poland}

\date{}
\begin{document}
\maketitle
\begin{abstract}

We discuss the chiral decomposition of non-symmetric stacking structures. It is shown that the low-energy electronic structure of Bernal stacked graphene multilayer deposited on h-BN consists of chiral pseudospin doublets. $N$-layer graphene stocks on h-BN layer have $N/2$ effective bilayer graphene systems and one effective h-BN layer if $N$ is even or $(N-1)/2$ effective graphene bilayers plus one graphene monolayer modified by h-BN layer if $N$ is odd. We present the decomposition procedure and we derive the recurrence relations for the effective parameters characterizing the chiral subsystems. The effective parameters consist in this case of the interlayer couplings and on-site potentials in contrast to pure graphene multilayer systems where only interlayer couplings are modified. We apply this procedure to discuss the Klein tunneling phenomena and compare quantitatively the results with pure graphene multilayer systems.

\end{abstract}

\newpage
\section{ Introduction}
Monolayer graphene consists of carbon atoms arranged in hexagonal structure which is not a Bravais lattice. It can be seen as a triangular lattice with a basis of two atoms per unit cell that leads, in reciprocal space, to a hexagonal Brillouin zone with Dirac cones in its corners. In contrast to the monolayer graphene, quasiparticles in bilayer graphene are massive chiral fermions due to its parabolic band structure. The perfect transmission in graphene monolayer and the perfect reflection in graphene bilayer for electrons incident in the normal direction at a potential barrier are viewed as two incarnations of the Klein paradox \cite{b1}.  The case of bilayer systems have been widely studied both theoretically \cite{b1} $\div$ \cite{b13} and experimentally \cite{b14} $\div$ \cite{b22}. For systems with three and more monolayers, electronic structures are also investigated theoretically \cite{b23} $\div$ \cite{b54} and experimentally \cite{b29} $\div$ \cite{b32} showing that multilayer graphene gives a number of intriguing properties including tunable band gap opening and anti-Klein tunneling, arising from chiral characteristics of charge carriers.\\
When $N$ layers are considered, the crystalline structure remains the same but now there are 2$N$ atoms per unit cell. One of the results concerning the multilayer systems gives proof of the variety of stacking structures. Graphene is usually produced by micromechanical cleavage of graphite so as a consequence the stacking structure is considered to be of graphite type. However, production of graphene with other stacking types is possible by, for example, epitaxial method. The band structures dependent on the stacking types are discussed \cite{b24}, \cite{b25} and the stability of different stacking systems is analysed in details \cite{b32}. However, theoretical and experimental works on chiral electron transport in multilayer systems have received less attention than in the case of mono-, bi- and tri-layer graphene. On the other hand, there is a generic interest in the possibility of engineering the electronic properties of two-dimensional crystals by combining them into multilayers \cite{b28}, \cite{b33} $\div$ \cite{b34}. In this context, it is interesting to see how the transport properties of graphene multilayers change by proximity to an insulator, so that the conducting channels would be only through graphene bands. Since graphene and hexagonal Boron Nitride (h-BN) have identical lattice structures, it is convenient to choose this insulator as a partner for heterogeneous multilayers \cite{b35}, \cite{b36}.\\
Theoretical studies of multilayer graphene revealed that Hamiltonian of Bernal stacking systems can be block diagonalized into effective bilayer (BLG) and monolayer (MLG) Hamiltonians depending on parity of layer numbers \cite{b26}, \cite{b37}. The properties of the effective bilayer subsystems are connected with chirality of charge carriers in each specific multilayer system. This result implies that it may be possible to find different chiral fermions in the multilayer graphene on h-BN systems to model tunneling properties in view of applications in nano-electronics. Although, h-BN is only weekly coupled with graphene and graphene bands sit deep within the hBN band gap, it is interesting to see which characteristics are really modified by its presence. In order to achieve this goal, the chiral decomposition procedure is generalized herein for multilayer graphene supported by h-BN layer. We turn our attention to the tunneling phenomena in such systems namely to the electron transport through the potential barrier higher than the incident electrons energy. We consider how the presence of h-BN layer appears in the electronic properties and in the possibility to observe interband tunneling when electrons outside the barrier (conduction band) transforms into holes inside it (valence band), or vice-versa \cite{b1}, \cite{b38}.

\section{Model considerations}

\begin{figure}
\includegraphics[scale=0.8]{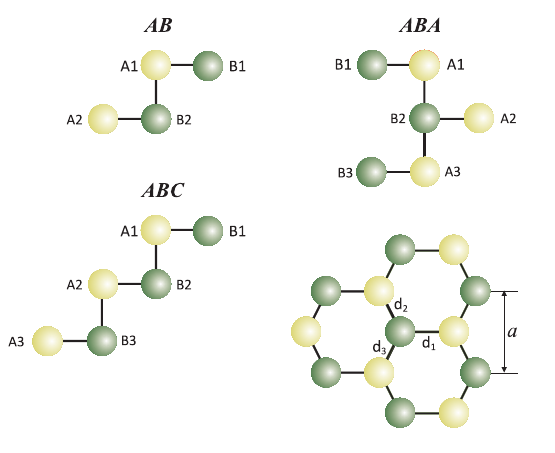}
\caption{\label{fig1}(Color online) Crystal structure of graphene (or h-BN) constructed by two interpenetrating triangular sublattices (yellow/lighter and green/darker balls). Side view of the unit cells for Bernal (ABA) and rhombohedral (ABC) stacking.}
\end{figure}

Layered materials are predominantly formed in hexagonal symmetries, including different stacking orders of the hexagonal layers composed of two triangular sublattices. According to Density Functional Theory (DFT) \cite{b39}, \cite{b40} and experimental findings \cite{b41}, it is energetically favorable for the atoms of sublattice A(B) to be displaced along the honeycomb edges in a way that an atom from the sublattice A(B) sits on top of an atom belonging to another sublattice B(A). This stacking rule implies the three distinct but equivalent projections of the 3D layered structures onto x-y plane and $ {2}^{N-2} $ distinct $N$-layer stack sequences \cite{b27}. It is also known that there are mainly two stacking types for graphite: the so-called Bernal stacking, forming the layer sequence 1212... and the rhombohedral stacking which form the layer sequence 123123... . These two different possibilities for $ N\leq 3 $ are shown in Fig.\ref{fig1}. 

The multilayer systems of these types can be described by the following tight-binding $\pi$ bands Hamiltonian:
\begin{equation}
\label{e1}
\begin{split}
H=H_{ML}+H_{inter}
\end{split}
\end{equation}
where two components account for the intralayer and interlayer hoppings, respectively. The trial wave function for such multilayer system can be written as:

\begin{equation}
\label{e2}
\begin{split}
\Psi_{\vec{k}}(\vec{r})=\sum_{j=1}^{2N}a_{\vec{k}}^{(j)}\Psi_{\vec{k}}^{(j)}(\vec{r})
\end{split}
\end{equation}
where the superscript $j$ denotes the different atoms per unit cell, $a_{\vec{k}}^{(j)}$ are complex functions of the quasi-momentum ${\vec{k}}$ and $ \Psi_{\vec{k}}^{(j)} $  are the Bloch functions constructed from the atomic orbital wave functions:
\begin{equation}
\label{e3}
\begin{split}
\Psi_{\vec{k}}^{(j)}(\vec{r})=\sum_{R_l}exp (i\vec{k}\cdot\vec{R_l})\phi^{(j)}(\vec{r}+\vec{\delta_j}-\vec{R_l})
\end{split}
\end{equation}
$\vec{\delta}_j $ is the vector which connects the sites of the underlying Bravais lattice with the site of the $j$ atom within the unit cell. There are $2N$ different energy bands and the energy of the $j$th band is given by   $E_{\vec{k}}^{(j)}=\langle\Psi_{\vec{k}}^{(j)} \vert H \vert \Psi_{\vec{k}}^{(j)}\rangle / \langle\Psi_{\vec{k}}^{(j)}\vert\Psi_{\vec{k}}^{(j)}\rangle$. Minimizing this energy with respect to the $a_{\vec{k}}^{(j)}$   coefficients leads to the eigenvalue equation $H\Psi_{\vec{k}}^{(j)} = E_{\vec{k}}^{(j)}S\Psi_{\vec{k}}^{(j)}$  . The transfer integral matrix $H$ and the overlap integral matrix $S$ are $2N\times2N$ matrices. The band energies may be determine from the eigenvalue equation by solving the secular equation $det(H-E_{\vec{k}}^{(j)}I)=0$.
\par
 We assume that the interaction between two atoms depends only on their distance and we take into account the nearest neighbours interactions. 
 \par
 The transfer integral matrix that takes into account different stacking sequences and on-site potential energies may be written as:

\begin{equation}
\label{e4}
H_N=\left(  
\begin{smallmatrix} 
\mathbf{ML}_1& \mathbf{\Gamma}_{1,2}&\\
\mathbf{\Gamma}_{2,1} & \mathbf{ML}_2&\mathbf{\Gamma}_{2,3}\\
       &\mathbf{\Gamma}_{3,2}&   \mathbf{ML}_3      &\mathbf{\Gamma}_{3,4}\\
       &      &\mathbf{\Gamma}_{4,3}&\cdots\\
     &  & &\ddots\\
     & & & & & \mathbf{ML}_{N-1} &\mathbf{\Gamma}_{N-1,N}\\
       & & & & &\mathbf{\Gamma}_{N,N-1}&\mathbf{ML}_N  
\end{smallmatrix}
\right) 
\end{equation}

\begin{equation}
\label{e5}
\mathbf{ML}_i=
\left(  
\begin{array}{cc}
  \varepsilon_{i,\alpha}&-\gamma_{0,i}f_i(\vec{k})\\
   -\gamma_{0,i}f_i^*(\vec{k})&\varepsilon_{i,\beta}
\end{array}
\right) \;\;\;\;\;i=1,2,....N
\end{equation}

\begin{equation}
\label{e6}
\mathbf{\Gamma}_{i,i+1}=\gamma_i
\left(  
\begin{array}{cc}
 0&s_i\\
   1-s_i&0 \\
\end{array}
\right) \;\;\;\;\;i=1,2,....N-1
\end{equation}

\begin{equation}
\label{e7}
\begin{split}
\mathbf{\Gamma}_{i+1,i}=\mathbf{\Gamma}^T_{i,i+1}
\end{split}
\end{equation}
Parameter $s_i$ is equal $0$ or $1$ depending on layers sequence in the multilayer system under consideration. The diagonal terms $  \varepsilon_{i,\alpha (\beta)}$  denote the on-site energy of electron at the atom in layer $i$ belonging to sublattice A(B). In the first approximation, they can be equal to the energy of an electron in the $2pz$ orbital of an atom. However, this energy is modified as atoms bond together forming the lattice and can be considered as a parameter to fit with the experimental findings. Two parameters $\gamma$   describe the strength of the coupling between a specific pair of atoms: $\gamma_{0,i}$ denotes the coupling between the nearest neighbors in each monolayer while $\gamma_i$  describes direct interlayer coupling. The geometrical factor $f(\vec{k})$  resulting from a summation over nearest neighbours in each monolayer can be written in terms of vectors $\vec{d_i} $ indicated in Fig.\ref{fig1}:

\begin{equation}
\label{e8}
\begin{split}
f(\vec{k})\equiv\sum_{i=1}^3exp(i\vec{k}\cdot\vec{d_i})=exp\left(i\frac{•k_ya}{\sqrt{3}}\right) +exp\left(i\frac{•k_ya}{2\sqrt{3}}\right)cos\left( \frac{k_xa}{2}\right) 
\end{split}
\end{equation}
and can be approximated by: $f(\vec{k})\approx-\frac{\sqrt{3}a}{2\hbar}(\xi p_x-ip_y)$ with  $\vec{p}=\hbar \vec{k}-\hbar\vec{K_\xi}, \xi\pm1$.

The overlap integrals take into account the fact that the orbitals do not span an orthogonal basis set. In the present approximation we include the overlap between two nearest neighbour atoms in monolayer and the overlap between the atoms which are directly above/below each other in neighbouring monolayers. Due to their small value, in all situations under consideration in this paper, even these two integrals are neglected. Then, we obtain the following recurrence relation for the determinant of the $N$-layer system:
\begin{equation}
\label{e9}
\begin{split}
&det(H_N-EI)=D_N=det \mathbf{ML}_N \cdot D_{N-1} \\
&+(\varepsilon_{N,\beta}-E)\left\lbrace \sum_{k=1}^{N-2}(-1)^k(\varepsilon_{N-k,\alpha}-E)\prod_{i=N-k}^{N-1} (1-s_i)^2\gamma_i^2D_{N-k-1}\right.\\
&\left.+(-1)^{N-1}(\varepsilon_{1,\alpha}-E)\prod_{i=1}^{N-1}(1-s_i)^2\gamma_i^2\right\rbrace \\
&+(\varepsilon_{N,\alpha}-E) \left\lbrace \sum_{k=1}^{N-2}(-1)^k(\varepsilon_{N-k,\beta}-E)\prod_{i=N-k}^{N-1}s_i^2\gamma_i^2D_{N-k-1}\right.\\
&\left.+(-1)^{N-1}(\varepsilon_{1,\beta}-E)\prod_{i=1}^{N-1}s_i^2\gamma_i^2 \right\rbrace 
\end{split}  
\end{equation}

\begin{figure}
\includegraphics[scale=0.8]{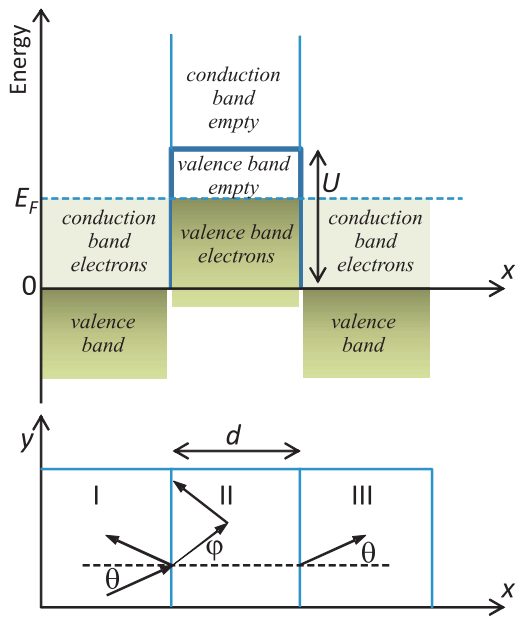}
\caption{\label{fig2}(Color online) Top panel: Schematic of the band structure across a potential barrier of height $U$ and width $d$. Bottom panel: Schematic of the potential barrier in the xy-plane. Charge carriers are incident from region I on the barrier (region II) and are transmitted in region III. The incident ($\theta$) and refractive ($\varphi$) angles are indicated, respectively.}
\end{figure}

Now, we focus on  the transport properties in such multilayer heterostructures. Let us consider the charge carriers incident on a potential barrier. Such a local potential barrier can be created, for example, by the electric field effect coming from local chemical doping. The situation is depicted schematically in Fig.\ref{fig2}. 

Particles travelling from the left (region I) to the right (region III) in graphene/hBN multilayer system are incident at energy $E_F>0$   on a potential barrier (region II) of height $U$  and width $d$ . \\

  The Schr\"odinger equation describing this situation can be written as:
\begin{align}
\label{e10}
&\left(  
\begin{smallmatrix}  
\mathbf{ML}_1& \mathbf{\Gamma}_{1,2}&\\
\mathbf{\Gamma}_{2,1} & \mathbf{ML}_2&\mathbf{\Gamma}_{2,3}\\
       &\mathbf{\Gamma}_{3,2}&   \mathbf{ML}_3      &\mathbf{\Gamma}_{3,4}\\
       &      &\mathbf{\Gamma}_{4,3}&\cdots\\
     &  & &\ddots\\
     & & & & & \mathbf{ML}_{N-1} &\mathbf{\Gamma}_{N-1,N}\\
       & & & & &\mathbf{\Gamma}_{N,N-1}&\mathbf{ML}_N   
\end{smallmatrix}
\right)
\left(
\begin{smallmatrix}
\Psi_1\\
 \Psi_2\\
 \Psi_3\\
 \vdots\\
 \vdots\\
 \vdots\\
 \Psi_N
 \end{smallmatrix}
 \right) 
 =
 &(E-U)\left(
\begin{smallmatrix}
\Psi_1\\
 \Psi_2\\
 \Psi_3\\
 \vdots\\
 \vdots\\
 \vdots\\
 \Psi_N
 \end{smallmatrix}
 \right) 
\end{align}
with
\begin{equation}
\label{e11}
\begin{split}
\Psi_i=
\left(  
\begin{array}{c}
 A_i\\
   B_i \\
\end{array}
\right) \;\;\;\;\;i=1,2,....N
\end{split}
\end{equation}
where $A_i$ and $B_i$ represent the wave functions of sublattice $A$ and $B$, respectively.

In order to find the eigenvectors  one has to solve the following set of recurrence equations:

\begin{align}
\label{e12}
&\delta_{1k}\left[  (\varepsilon_{1,\alpha}-E)A_1-\gamma_{0,1}f_1(\vec{k})B_1+s_1\gamma_1B_2\right]  \nonumber \\ 
&+(1-\delta_{1k})(1-\delta_{Nk})
\left[ (1-s_k)\gamma_kB_{k-1} + (\varepsilon_{k,\alpha}-E)A_k-\gamma_{0,k}f_k^*(\vec{k})B_k+s_k\gamma_kB_{k+1}\right]  \nonumber \\
&+\delta_{Nk}\left[ (1-s_N)\gamma_NB_{N-1} + (\varepsilon_{N,\alpha}-E)A_N-\gamma_{0,N}f_N(\vec{k})B_N\right] =0 \nonumber \\
&\delta_{1k}\left[  -\gamma_{0,1}f_1^*(\vec{k})A_1+(\varepsilon_{1,\beta}-E)B_1+(1-s_1)\gamma_1A_2)+
\right]   \nonumber \\
&+(1-\delta_{1k})(1-\delta_{Nk})
\left[ (s_k\gamma_kA_{k-1} + (\varepsilon_{k,\beta}-E)B_k-\gamma_{0,k}f_k^*(\vec{k})A_k+(1-s_k)\gamma_kA_{k+1}\right]  \nonumber \\
&+\delta_{Nk}\left[ s_N\gamma_NA_{N-1} + (\varepsilon_{N,\beta}-E)B_N-\gamma_{0,N}f_N^*(\vec{k})A_N\right] =0 
\end{align}
for $k=1,2,...N$.\\

The plane-wave solution can be then written as $2N$ spinor:
\begin{equation}
\label{e13}
\begin{split}
\Psi_R=\left( c_1^R\Psi_{k_x}e^{-ik_xx}+ c_2^R\Psi_{-k_x}e^{ik_xx}  + c_3^R\Psi_{i\kappa_x}e^{\kappa_xx}            + c_4^R\Psi_{-i\kappa_x}e^{-\kappa_xx} \right)e^{ik_yy} 
\end{split}  
\end{equation}
where index $R$ represents three regions: $R=I$ on the left of barrier ($c^I_4=0$), $R=II$  inside the barrier and  $R=III$ on the right of barrier ($c^{III}_2=c^{III}_3=0$). Let $k^2=2mE_F/\hbar^2$  be the wave vector for propagating modes in the regions $I$ and $III$, while $k^2=2m|E_F-U|/\hbar^2$   is the wave vector in region II. We define then $k_x=k cos\vartheta$  and $k_y= k sin\vartheta$ in I and III regions while in region II we have  $k_x=k cos\varphi$  and $k_y= k sin\varphi=ksin\vartheta$  , the last because the transverse momentum is conserved. To be able to fulfill the boundary conditions for tunneling through the potential barrier we have to consider the decaying modes which have an imaginary value of the momentum in the x-direction: $k_x=i\kappa_x$  . Now the coefficients $c^R_{1,2,3,4}$  have to be found from the continuity of $\Psi_R$  at the boundaries between regions $I$ and $II$ ($x=0$) and $II$ and $III$ ($x=d$). It is well known that the most interesting behaviour is expected for $U>E_F$, were the chiral tunneling can be observed. All results in the following are discussed in this context.

\section{Results and discussion }

As described in Section 2, we assume the nearest neighbors interactions and not locally modified onsite energies. It means we neglect the asymmetry leading to "mexican hat" in band structure, the trigonal warping and opening up of small energy gaps between the conduction and the valence bands. The details of the band structure including all these effects is a very complicated problem that has yet to be studied. In view of the large uncertainty in the parameters involved, it is meaningless to introduce, at this stage, a more complicated model. However, it should be underline that inclusion of the other parameters does not cause principal problems, but the analysis becomes more complicated. For example, the inclusion of trigonal warping phenomenon does not block the possibility of chiral decomposition of the Hamiltonian describing the graphene multilayer systems with arbitrary number of layers \cite{b34}. Hence, we believe that the Hamiltonian (\ref{e4}) correctly captures the main features of the present problem. 
\par
The values of all parameters employed in the present numerical calculations are taken from the standard set of parameters used in the theoretical analysis of the graphene/h-BN multilayer systems properties \cite{b42}, \cite{b43}. Within the error bar they agree with the experimentally estimated parameters \cite{b44}. Two parameters describing the strength of the coupling between a specific pair of atoms are taken to be: $\gamma_{0,iC-C}\equiv\gamma_0=3,033$ eV  and $\gamma_{0,iN-B}=2,36$ eV   for coupling between the nearest neighbours in each graphene and h-BN monolayer, respectively while $\gamma_{\mathrm{i,C-C}} \equiv\gamma_1=0,39$ eV  and $\gamma_{\mathrm{i,C-N}}=\gamma_{\mathrm{1,C-N}}=0,25$ eV describe interlayer couplings. Note, that we consider only stockings with Carbon-Nitrogen interaction \cite{b42}. Moreover, it is assumed here that lattice constant $a$  is common for all monolayers building a system. Thus, the difference between lattice constant of graphene and h-BN is not taken into account. However, it seems to be justified by the fact that we consider the ultrathin multilayer systems which are willing to accommodate \cite{b45} in contrast to graphene on h-BN substrate when moir\'e pattern is observed \cite{b35}. Furthermore, we take the onsite energy to be equal zero for Carbon, $3.36$ eV for  Nitrogen and $-3.66$ eV for Boron \cite{b42}. 
\par
Transmission probability through a $100$ nm-wide barrier is calculated as a function of incident angle for the Fermi energy of incident electrons of $17$ meV which is most typical in experiments with graphene \cite{b29}, \cite{b46}. The barrier height is taken to be $50$ meV. For better comparison, the parameters are kept the same for all considered systems unless other values are stated in the text.

\subsection{Homogenous multilayers }
Although, the main results of the present paper are connected with graphene/h-BN hybrid systems, we introduce likewise some remarks on pure graphene multilayers which fits in the mainstream of discussion about tunneling effects (\cite{b52}, \cite{b53} and references in there) and seems to be interesting for better understanding the interband tunneling nuances. The results of this subsection serve simultaneously as a background for comparison with the results for heterogeneous multilayers.

\begin{figure}
\includegraphics[scale=0.5]{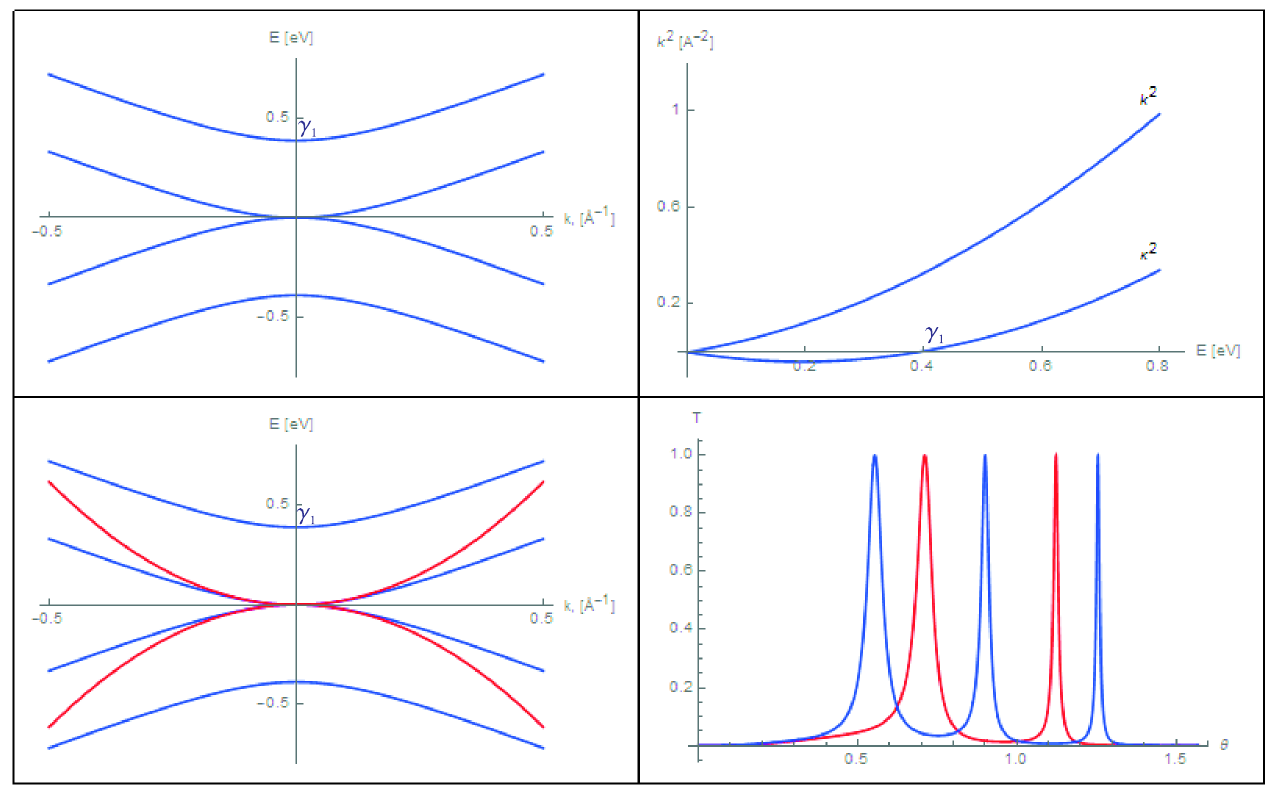}
\caption{\label{fig3}(Color online) Top panel: Band structure of BLG and corresponding dependence of absolute momentum on energy. Bottom panel: Band structure of BLG in four- (blue curves) and two-band (red curves) approach with corresponding transmission probabilities as a function of incident angle.}
\end{figure}

Let us start with a few comments on bilayer graphene (BLG). Eigenvalues of the Hamiltonian (\ref{e4}) for bilayer graphene are given by:
\begin{equation}
\label{e14}
\begin{split}
E^\eta_\pm= \pm\frac{\gamma_1}{2} \left(  \sqrt{1+\frac{4\gamma_0^2\mid f(\vec{k})\mid^2}{\gamma_1^2}}+\eta  \right), \;\;\;\eta=\pm1
\end{split}
\end{equation}
The eigenvalues$E^+_\pm$  describe two bands that are splitting away from zero energy by $\pm\gamma_1$ at the K point ($|f(k)|=0$). This is because the orbitals on the $A2$ and $B1$ sites form a dimer that is coupled by interlayer hopping $\gamma_1$, resulting in a bonding and anti-bonding pair of states $\pm\gamma_1$.
The formula $E^-_\pm$  interpolates between linear dispersion at large momenta ($\gamma_1<<\frac{\sqrt{3}a\gamma_0}{2\hbar}p<\gamma_0$) and quadratic dispersion near zero energy where the bands touch. These bands arise from effective coupling between the orbitals on sites $A1$ and $B2$, that don't have a counterpart in the other layer. Such a system represents a gapless semiconductor with chiral electrons and holes with a finite mass. The bilayer graphene Hamiltonian, written in a two-component basis \cite{b1}, \cite{b4}, yields a parabolic energy spectrum and it is straightforward that we have four possible solutions for a given energy. Two of them correspond to propagating waves and the other two to evanescent waves. It leads to an intriguing behaviour in transport through the potential barrier higher than the incident electrons energy: electrons outside the barrier transforms into holes inside it, or vice-versa \cite{b1}. The limit of the validity for the two band model lies in the condition that two high energy bands are not occupied which seems to be fulfilled in BLG. However, it is interesting to see that this intriguing behavior can be observed also in the four bands description of BLG. In the top panel of Fig.\ref{fig3} we show the band structure and the corresponding dependence of absolute momentum $k^2$ on energy $E$. 

One can notice that in the energy range $0<E<\gamma_1$  we still have two traveling waves  and two evanescent waves $k^2=\kappa^2<0$ which have an imaginary value of momentum in the propagation direction (holes with wave vectors $i\kappa$). We can consider now the tunneling phenomena in four bands approach. We assume that   $0<E_F<\gamma_1$ for the incoming wave as this is likely the experimental situation. It is straightforward that the imaginary value of the momentum in the x-direction can be written as: $\kappa_x=\sqrt{k_y-\kappa^2}$  and  in the case of BLG it takes a form: $\kappa_x=\sqrt{k_y^2-[4/\alpha (1-\sqrt{1+\alpha k^2})+k^2]}$ with $\alpha=3\gamma_0^2a^2/\gamma_1^2$  which reduces to $\kappa_x=\sqrt{2k_y^2+k_x^2}$ in two bands approximation. The dispersion relations in both approaches are compared in the bottom panel of Fig.\ref{fig3}. The corresponding transmission probabilities (Fig.\ref{fig3}-bottom panel right) exhibit "magic angles" in the spectrum, at which the total transmission is observed, however these angles differ between these two approaches. 
\par

\begin{figure}
\includegraphics[scale=0.5]{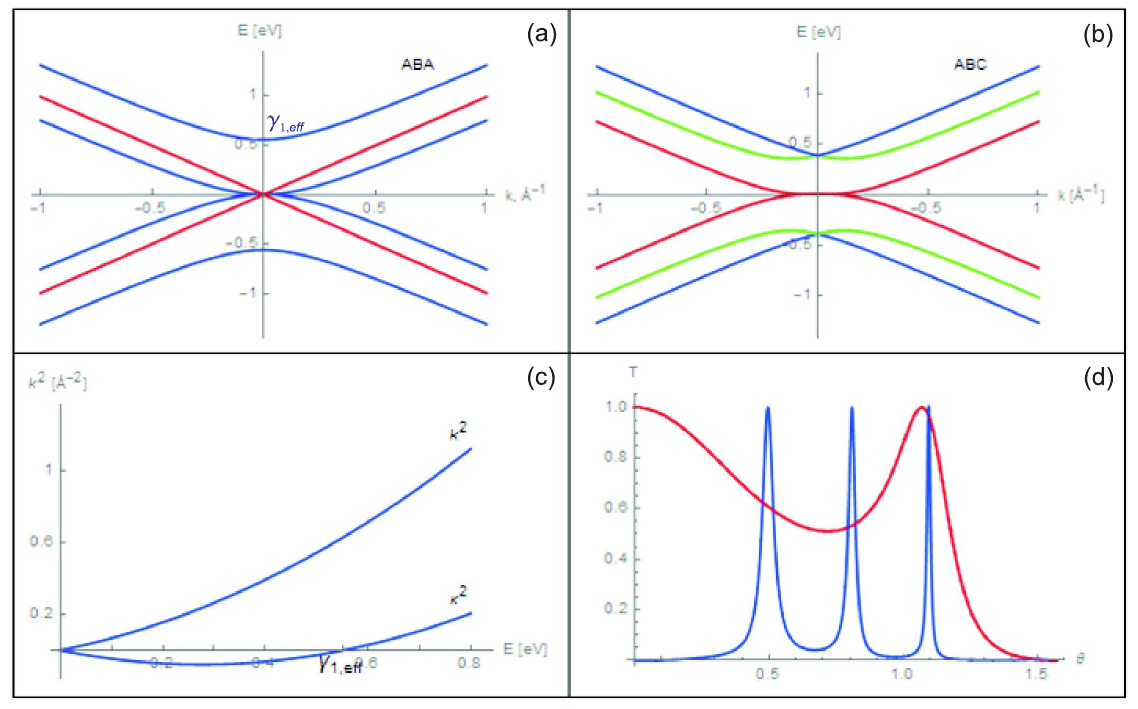}
\caption{\label{fig4}(Color online) Band structure of TLG stacked as (a) ABA (Bernal) and (b) ABC (rhombohedral). (c) Absolute momentum as a function of energy for effective BLG from TLG with Bernal stacking. (d) Transmission probability of Bernal stacked TLG as a function of incident angle with two modes: for effective BLG and MLG.}
\end{figure}

In order to calculate physical quantities of multilayer graphene system, it is useful to follow the partition procedure described in paper \cite{b27}. This procedure is based on division into smaller multilayer elements in the manner that excludes layers contained within previously identified partitions. It follows that there is only one way of stocking allowing the chiral decomposition into isolated bilayer systems with some effective interlayer hopping plus one monolayer if number of monolayers $N$ is odd. As an example, we present in Fig.\ref{fig4}a,b the band structure of trilayer graphene (TLG) stacked as ABA (Bernal) and ABC, respectively.

Only in the case of ABA stacking, the transfer integral matrix:

\begin{align}
\label{e15}
&H_{TLG}=     \nonumber \\
&\left(  
\begin{array}{cccccc}  
0 & -\gamma_0f(\vec{k}) & 0& \gamma_1 & 0 & 0    \\
-\gamma_0f^*(\vec{k}) & 0 & 0 & 0 & 0 & 0  \\
 0 & 0 & 0 & -\gamma_0f(\vec{k}) & 0 & 0  \\                                    
 \gamma_1 & 0 & -\gamma_0f^*(\vec{k}) & 0 & \gamma_1 & 0  \\
  0 & 0 & 0 & \gamma_1 & 0 & -\gamma_0f(\vec{k}) \\                                                   
  0 & 0 & 0 & 0 & -\gamma_0f^*(\vec{k}) & 0  
\end{array}
\right) 
\end{align}
can be rewritten in terms of two blocks: one represents an effective BLG and another one describes usual graphene monolayer (MLG):
\begin{align}
\label{e16}
&H_{TLG}^{eff}=     \nonumber \\
&\left(  
\begin{array}{cccccc}  
0 & -\gamma_0f(\vec{k}) & 0& \gamma_{1,eff} & 0 & 0    \\
-\gamma_0f^*(\vec{k}) & 0 & 0 & 0 & 0 & 0  \\
 0 & 0 & 0 & -\gamma_0f(\vec{k}) & 0 & 0  \\                                    
 \gamma_{1,eff} & 0 & -\gamma_0f^*(\vec{k}) & 0 & 0 & 0  \\
  0 & 0 & 0 & 0 & 0 & -\gamma_0f(\vec{k}) \\                                                   
  0 & 0 & 0 & 0 & -\gamma_0f^*(\vec{k}) & 0  
\end{array}
\right) 
\end{align}
where $\gamma_{1,eff}=\sqrt{2}\gamma_1$ \cite{b26}.
\par
As a consequence, in the energy range $0<E<\gamma_{1,eff}$  we again have two traveling waves $k^2>0$ and two evanescent waves $k^2\equiv \kappa^2<0$  , which have an imaginary value of momentum in the propagation direction (see Fig.\ref{fig4}c). The transmission probability of such TLG has two modes: one connected with effective BLG and the second one connected with MLG. Both modes are shown in Fig.\ref{fig4}d. It is easily seen that for certain incident angles one can expect the tunneling through the potential barrier coming both from BLG and MLG. As for ABC stacked trilayer, we have complex solutions for $k^2$  in the energy range $0<E<\gamma_1$  with no clear physical interpretation, in contrast to previous cases when we have to do with holelike states \cite{b1}. Similar conclusions arise also from Ref. \cite{b26}, \cite{b37}. Therefore, calculations in the following are mainly devoted to Bernal stacking systems. 

\subsection{Heterogeneous multilayers}

We consider now, the problem of chiral decomposition in graphene multilayers supported by  h-BN layer. First of all, let us recall the electronic properties of graphene/h-BN bilayer system. Due to the both tight-binding \cite{b47} and DFT calculations \cite{b48}, its band structure shows a gap between valence and conduction bands. However, in the vicinity of the Brillouin-zon corners the low-energy band structure possesses nearly graphene character indicating a weak interaction between the layers. This last observation has turned our attention towards the systems with more graphene layers deposited on h-BN layer which could be viewed as support protecting the graphene properties. The simplest system where one could expect chiral decomposition can consists of 3ML, two graphene on one h-BN.
The Hamiltonian matrix (\ref{e4}) for BLG/h-BN system takes the form:

\begin{align}
\label{e17}
&H_{BLG/h-BN}=     \nonumber \\
&\left(  
\begin{array}{cccccc}  
0 & -\gamma_0f(\vec{k}) & 0& \gamma_1 & 0 & 0   \\
-\gamma_0f^*(\vec{k}) & 0 & 0 & 0 & 0 & 0 \\
 0 & 0 & 0 & -\gamma_0f(\vec{k}) & 0 & 0  \\                                    
 \gamma_1 & 0 & -\gamma_0f^*(\vec{k}) & 0 & \gamma_{\mathrm{1,C-N}}  & 0  \\
  0 & 0 & 0 &  \gamma_{\mathrm{1,C-N}}  & \varepsilon_{\mathrm{N}} & -\gamma_0f(\vec{k})  \\                                                   
  0 & 0 & 0 & 0 & -\gamma_0f^*(\vec{k}) & \varepsilon_{\mathrm{B}} \\
  \end{array}
\right)  
\end{align}
and the effective Hamiltonian, with the same eigenvalues as the original one shown above, can be written as:

\begin{align}
\label{e18}
&H_{BLG/h-BN}^{eff}=     \nonumber \\
&\left(  
\begin{array}{cccccc}  
\varepsilon & -\gamma_0f(\vec{k}) & 0& \gamma_{1,eff} & 0 & 0   \\
-\gamma_0f^*(\vec{k}) & 0 & 0 & 0 & 0 & 0 \\
 0 & 0 & 0 & -\gamma_0f(\vec{k}) & 0 & 0  \\                                    
 \gamma_{1,eff} & 0 & -\gamma_0f^*(\vec{k}) & \varepsilon& 0 & 0  \\
  0 & 0 & 0 &0 & \varepsilon'_{\mathrm{N}} & -\gamma_0f(\vec{k})  \\                                                   
  0 & 0 & 0 & 0 & -\gamma_0f^*(\vec{k}) & \varepsilon_{\mathrm{B}} \\
  \end{array}
\right)  
\end{align}
The effective Hamiltonian is derived in the same way as for multilayer graphene systems \cite{b26}, it means by assuming $det(H-EI)=det(H^{eff}-EI)$ . In the present case, however the effective Hamiltonian is more complicated. It contains not only the effective hoppings but also the effective onsite energies both for bilayer graphene and h-BN layer. These effective parameters are related to the original parameters by formulas:

\begin{equation}
\label{e19}
\begin{split}
&\varepsilon_{\mathrm{N}}=\varepsilon'_{\mathrm{N}}+2\varepsilon \\
&\gamma_1^2+\gamma^2_{1,C-N}=-2\varepsilon\varepsilon'_{\mathrm{N}}-(\varepsilon^2-\gamma^2_{1,eff}) \\
&-\varepsilon_{\mathrm{N}}\gamma_1^2=\varepsilon'_{\mathrm{N}}(\varepsilon^2-\gamma^2_{1,eff})
\end{split}
\end{equation}
coming from the factorization procedure applied to the determinant (\ref{e9}). Solving numerically the set of equations (\ref{e19}), all effective parameters can be found and for the present case they take the values: $\gamma_{1,eff}=0.389$ eV, $\varepsilon=-0.0094$ eV and $\varepsilon'_{\mathrm{N}}=3.3766$ eV.

\begin{figure}
\includegraphics[scale=0.5]{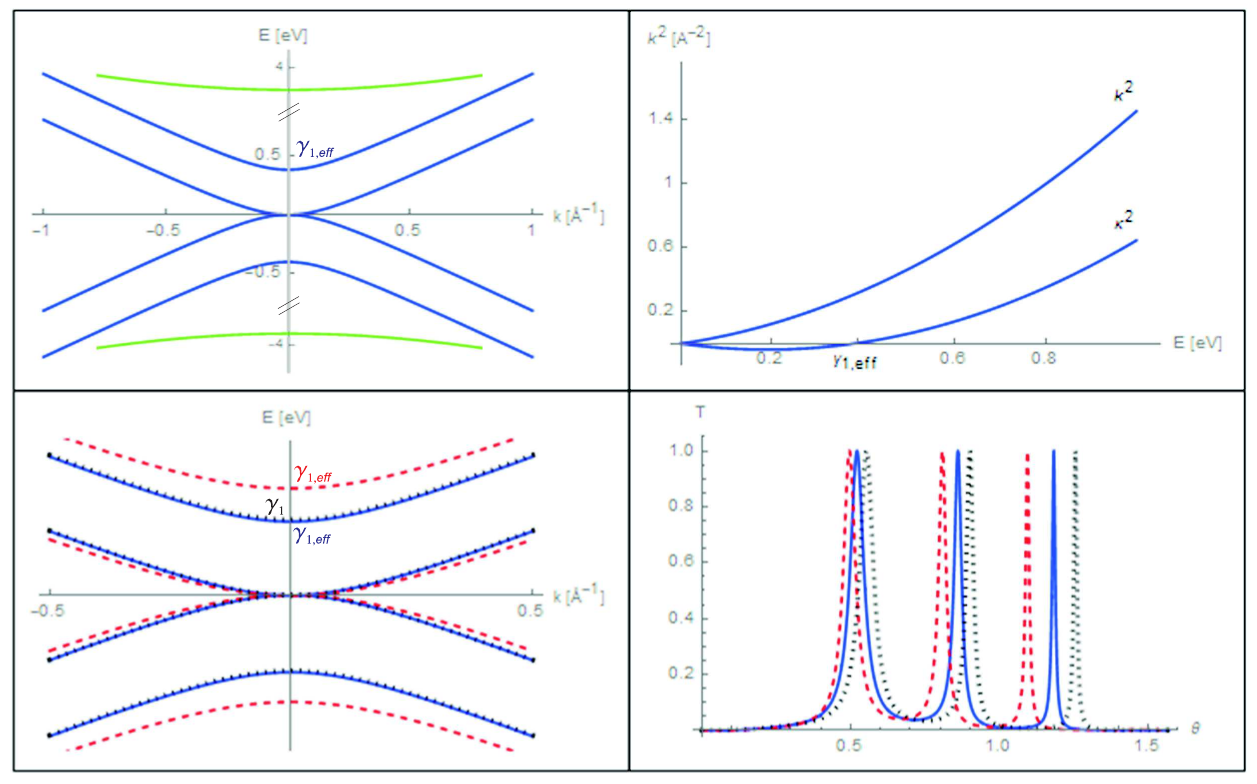}
\caption{\label{fig5}(Color online) Top panel: Band structure of BLG/h-BN and corresponding dependence of absolute momentum on energy.  Bottom panel left: Band structure of free-standing BLG (black dotted curves), effective BLG from TLG (red dashed curves) and effective BLG from BLG/h-BN (blue solid curves). Bottom panel right: Transmission probability as a function of incident angle for free-standing BLG (black dotted curve), effective BLG from TLG (red dashed curve) and effective BLG from BLG/h-BN (blue solid curve).}
\end{figure}

In top panel of Fig.\ref{fig5}, we present the band structure of BLG/h-BN together with the corresponding dependence of absolute momentum $k^2$ on energy $E$. A simple inspection of these curves indicates that we have to do with one effective BLG and single h-BN layer wherein the band structure of this BLG differ from the band structures of free-standing BLG and effective BLG in trilayer graphene system (see Fig.\ref{fig5} bottom panel left). In consequence, the transmission probability is also different for these three kind of BLG. The transmission probability presented in Fig.\ref{fig5} (bottom panel right) exhibits perfect tunnelling at some incident angles characteristic for a given system. It is worth noting that the differences in electronic properties of these three BLGs are pronounced better in Klein paradox phenomena than in the dispersion spectra. As it could be expected, the h-BN induced more subtle changes than modifications due to the third graphene monolayer. 

It is worth to notice here, that the effective on-site potential $\varepsilon$ induced in graphene layers does not give rise to the gap in the spectrum. It acts rather as a compensation against the imposed asymmetry by h-BN support. 

\begin{figure}
\includegraphics[scale=0.5]{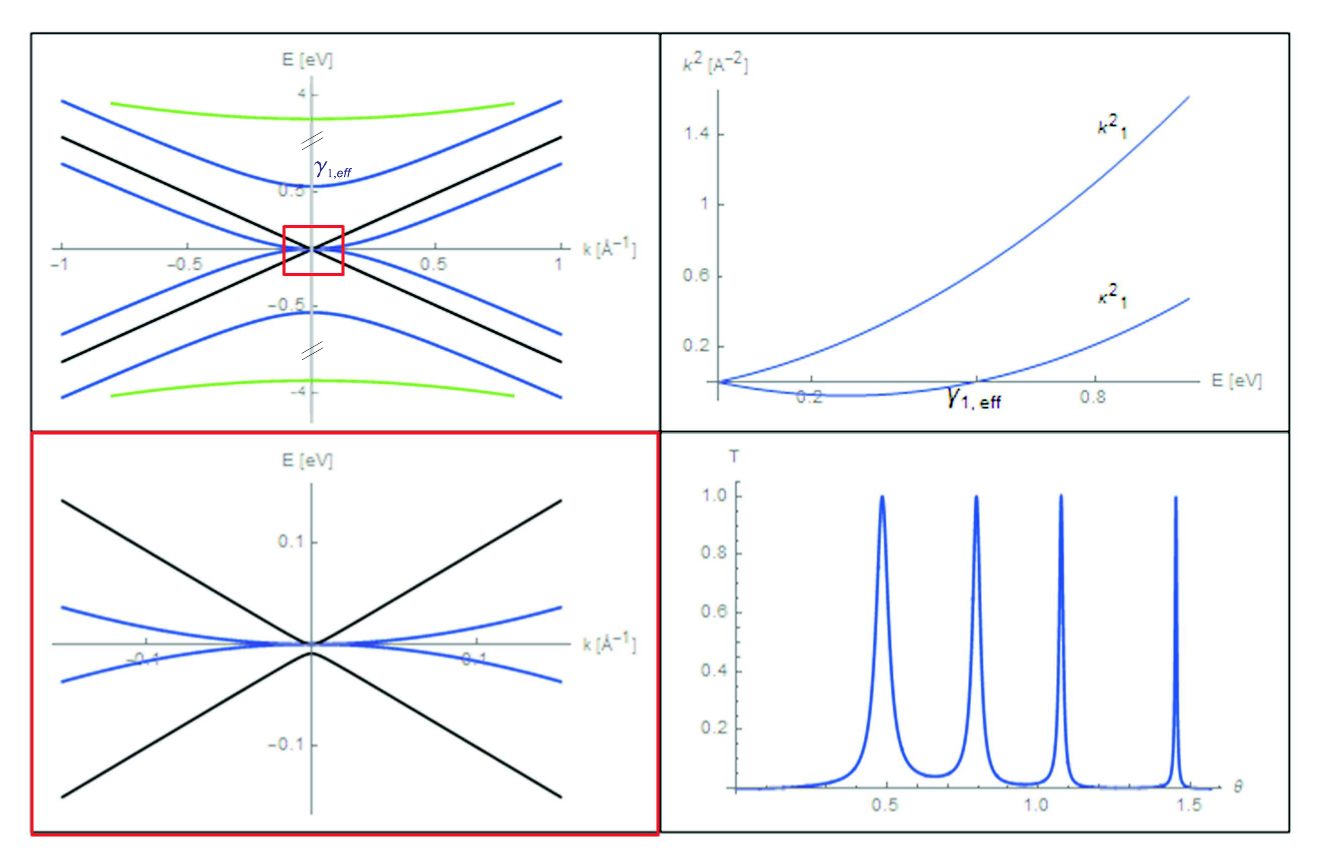}
\caption{\label{fig6}(Color online) Top panel: Band structure of TLG/h-BN and corresponding dependence of absolute momentum on energy. Bottom panel left: Inset presenting window drawn on the dispersion picture. A visible band gap is about 9 meV. Bottom panel right: Transmission probability as a function of incident angle in effective BLG of TLG/h-BN system.}
\end{figure}

Next, we consider the system composed of three graphene layers and one h-BN layer for which the Hamiltonian matrix is given by:
\begin{align}
\label{e20}
&H_{TLG/h-BN}=    \nonumber  \\
&\left(  
\begin{smallmatrix}  
0 & -\gamma_0f(\vec{k}) & 0& \gamma_1 & 0 & 0 & 0 & 0  \\
-\gamma_0f^*(\vec{k}) & 0 & 0 & 0 & 0 & 0 & 0 & 0  \\
 0 & 0 & 0 & -\gamma_0f(\vec{k}) & 0  & 0 & 0 & 0  \\                                    
 \gamma_1 & 0  & -\gamma_0f^*(\vec{k}) & 0 & \gamma & 0 & 0  & 0  \\
  0&  0  & 0 & \gamma & 0  & -\gamma_0f(\vec{k})  & 0 & \gamma_{\mathrm{1,C-N}}  \\
0 & 0 & 0 & 0 & -\gamma_0f^*(\vec{k}) & 0 & 0 & 0   \\
0& 0 &  0 & 0 & 0 &0 & \varepsilon_{\mathrm{B}} & -\gamma_0f(\vec{k})  \\                                                   
0 & 0 & 0 & 0 & \gamma_{\mathrm{1,C-N}}  & 0 & -\gamma_0f^*(\vec{k}) & \varepsilon_{\mathrm{N}}  \\
  \end{smallmatrix}
\right)  
\end{align}
In top panel of Fig.\ref{fig6}, the band structure of TLG/h-BN system and corresponding dependence of absolute momentum $k^2$ on energy  $E$ are shown. 

The BLG bands can be immediately recognized in the spectrum. But, in contrast to homogenous TLG, the graphene monolayer does not conserve its properties. It behave similarly as in the graphene/h-BN system (see inset in bottom panel of Fig.\ref{fig6}), however, the band gap is twice smaller than in graphene/h-BN bilayer. 
The electronic properties of bilayer graphene in this system are different from the previous cases. The effective Hamiltonian takes the form:

\begin{align}
\label{e21}
&H_{TLG/h-BN}^{eff}=     \nonumber \\
&\left(  
\begin{smallmatrix}  
\varepsilon & -\gamma_0f(\vec{k}) & 0& \gamma_{1,eff} & 0 & 0 & 0 & 0  \\
-\gamma_0f^*(\vec{k}) & 0 & 0 & 0 & 0 & 0 & 0 & 0  \\
 0 & 0 & 0 & -\gamma_0f(\vec{k}) & 0  & 0 & 0 & 0  \\                                    
 \gamma_{1,eff} & 0  & -\gamma_0f^*(\vec{k}) & \varepsilon & 0 & 0 & 0  & 0  \\
  0&  0 & 0& 0 & 0  & -\gamma_0f(\vec{k})  & 0 & \gamma_{1,\mathrm{C-N},eff}  \\
0 & 0 & 0 & 0 & -\gamma_0f^*(\vec{k}) & 0 & 0 & 0   \\
0& 0 &  0 & 0 & 0 &0 & \varepsilon_{\mathrm{B}} & -\gamma_0f(\vec{k})  \\                                                   
0 & 0 & 0 & 0 & \gamma_{1,\mathrm{C-N},eff} & 0 & -\gamma_0f^*(\vec{k}) & \varepsilon'_{\mathrm{N}}  \\
  \end{smallmatrix}
\right)  
\end{align}
with the following dependences between effective and original parameters, hopping and onsite energies, respectively:
\begin{equation}
\label{e22}
\begin{split}
&\varepsilon'_{\mathrm{N}}= \varepsilon_{\mathrm{N}}- 2\varepsilon \\
&2\gamma^2_1+\gamma^2_{1,\mathrm{C-N}}=-2\varepsilon'_{\mathrm{N}}\varepsilon-(\varepsilon^2-\gamma^2_{1,eff})+ \gamma^2_{1,\mathrm{C-N},eff} \\
& -2\varepsilon_{\mathrm{N}}\gamma^2_1=\varepsilon'_{\mathrm{N}}(\varepsilon^2-\gamma^2_{1,eff})-2\varepsilon\gamma^2_{1,\mathrm{C-N},eff}\\
&-\gamma^2_{1,\mathrm{C-N}}\gamma^2_1=\gamma^2_{1,\mathrm{C-N},eff}(\varepsilon^2-\gamma^2_{1,eff})
\end{split}
\end{equation}

\par
The following values of effective parameters: $\gamma_{1,eff}=0.5509$ eV, $\varepsilon=-0.0047$ eV, $\varepsilon'_{\mathrm{N}}=3.3674$ eV, $\gamma_{1,\mathrm{C-N},eff}=0.177$ eV are found for the considered system. Bottom panel of Fig.\ref{fig6} shows the transmission probability for the effective BLG in TLG/h-BN system. 

\par
In general, the $N$-layer graphene Bernal stacking system deposited on h-BN layer can be described by $N/2$ isolated bilayer systems with some effective interlayer hopping and onsite energies and one h-BN with effective onsite energy if $N$ is even or $(N-1)/2$ bilayers plus one MLG modified by h-BN layer if $N$ is odd. In these last cases the MLG/h-BN bilayer is characterized by the effective interlayer hopping and effective onsite energy in h-BN sublattice which interact with appropriate MLG sublattice. The appropriate in the sense that the Bernal stacking is conserved in the system. This is exact mapping of equation (\ref{e4}) without using any approximation during the chiral decomposition procedure. We can easily obtain the recurrence relations for the effective parameters.\\  
If the number of graphene layers is even, $N=2N_0$,§ we have:

\begin{align}
\label{e23}
&\left.
\begin{array}{ll}
\varepsilon_{\mathrm{N}}B^{2N_0}_l= \varepsilon'_{\mathrm{N}}A^{2N_0}_{2l}-A^{2N_0}_{2l+1}   \\
\gamma^2_{1,\mathrm{C-N}}B^{2N_0-1}_l-B^{2N_0}_{l+1}=\varepsilon'_{\mathrm{N}}A^{2N_0}_{2l+1}-A^{2N_0}_{2l+2}
\end{array}
\right\}
\text{for } l=0,1,2...N_0-1 \\
&\;\;\;\varepsilon_{\mathrm{N}}B^{2N_0}_{N_0}=\varepsilon'_{\mathrm{N}}A^{2N_0}_{2N_0}& \nonumber   
\end{align}
If the number of graphene layers is odd $N=2N_0+1$ we have:

\begin{align}
\label{e24}
&\varepsilon_{\mathrm{N}}B^{2N_0+1}_l= \varepsilon'_{\mathrm{N}}A^{2N_0}_{2l}-A^{2N_0}_{2l+1}+
\gamma^2_{1,\mathrm{C-N},eff}A^{2N_0}_{2l-1} &\text{for } l&=1,...N_0-1  \nonumber\\
&\gamma^2_{1,\mathrm{C-N}}B^{2N_0}_l-B^{2N_0+1}_{l+1}=\varepsilon'_{\mathrm{N}}A^{2N_0}_{2l+1}-A^{2N_0}_{2l+2}
+\gamma^2_{1,\mathrm{C-N},eff}A^{2N_0}_{2l} &\text{for } l&=0,...N_0-1 \nonumber\\
&\varepsilon_{\mathrm{N}}B^{2N_0+1}_{N_0}=\varepsilon'_{\mathrm{N}}A^{2N_0}_{2N_0} +\gamma^2_{1,\mathrm{C-N},eff}A^{2N_0}_{2N_0-1}  \\
&\gamma^2_{1,\mathrm{C-N}}B^{2N_0}_{N_0}=\gamma^2_{1,\mathrm{C-N},eff}A^{2N_0}_{2N_0} \nonumber \\
&\varepsilon_{\mathrm{N}}=\varepsilon'_{\mathrm{N}}-A^{2N_0}_{1} \nonumber
\end{align}
In the both above relations we introduced the following notation:
\begin{align}
\label{e25}
B^N_l&=\left\{
\begin{array}{ll}
\sum\limits_{P^{[N/2]}_l}\quad \prod \limits_{i \in P^{[N/2]}_l}(-\lambda^2_{N,N+1-2i}) & \text{if}\  l>0 \nonumber\\
1 & \text{if}\  l=0
\end{array} \right. \\
\lambda_{N_1,N_2}&=2\gamma sin \frac{N_2}{2(N_1+1)}\pi  \nonumber \\
A^{2N_0}&=
\left\{
\begin{array}{ll}
\sum\limits_{P^{2N_0}_l}\quad \prod \limits_{i \in P^{2N_0}_l}(-\lambda_i) &\text{if}\ \ l>0 \\
1&\text{if}\ \ l=0
\end{array} \right. \\
\lambda_i&=
\left\{
\begin{array}{ll}
\varepsilon_i-\gamma_{i,C-C,eff}  &\text{if} \ \ i=2k \nonumber\\
\varepsilon_i+\gamma_{i,C-C,eff}  &\text{if} \ \ i=2k+1
\end{array} \right. 
\end{align}
and $P^N_l$ denotes an $l$-elements subset of the set $\{1,2,....N\}$ and $\sum_{P^N_l}$ means the sum over all $l$-elements subsets.

\begin{figure}
\includegraphics[scale=0.5]{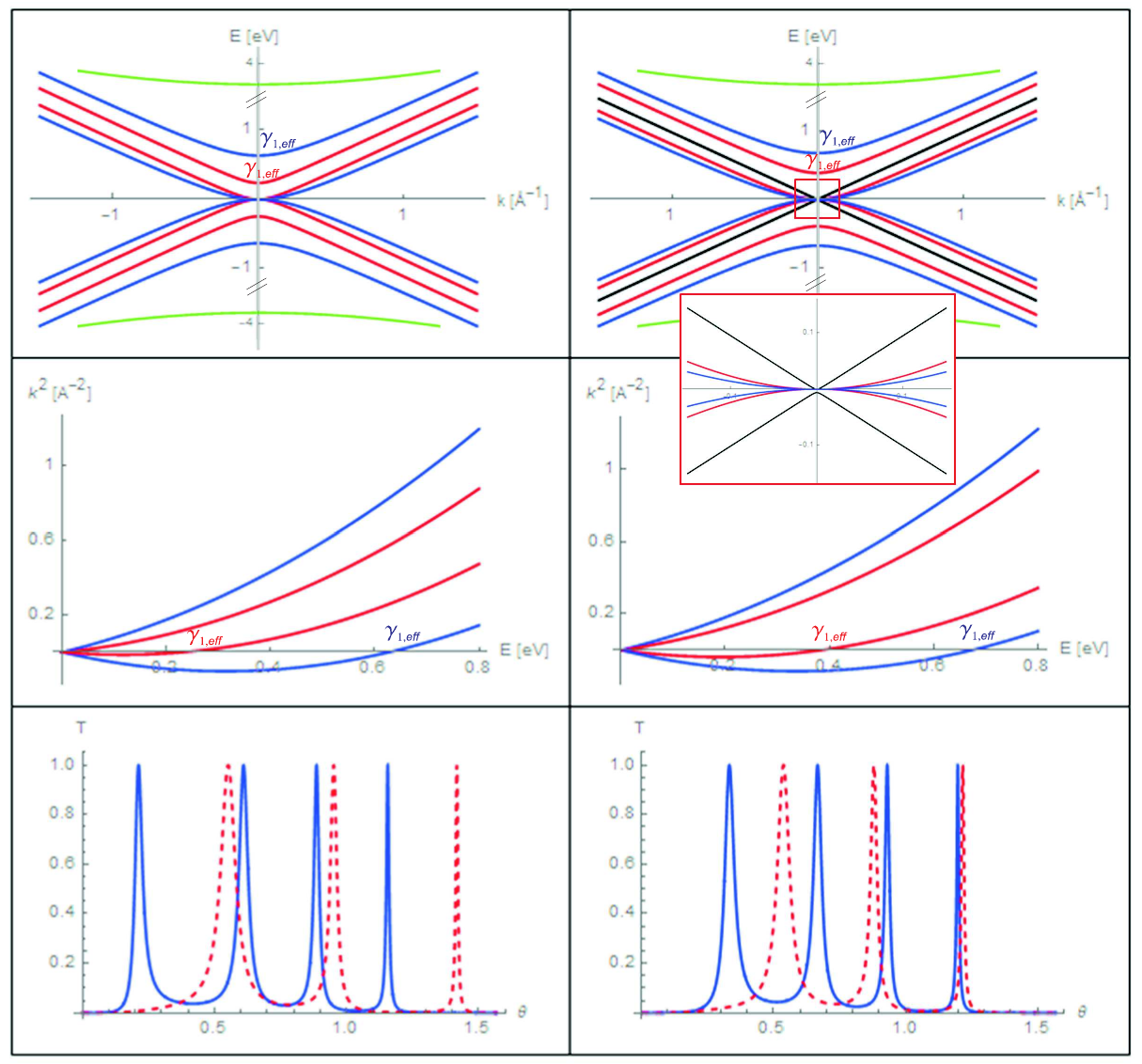}
\caption{\label{fig7}(Color online) Band structure, absolute momentum as a function of energy and transmission probability as a function of incident angle for multilayer graphene composed of 4 and 5 MLG deposited on one h-BN layer.Band gap visible in the inset is about 6 meV.}
\end{figure}

In Fig.\ref{fig7}, we collected the results for multilayer systems with $N=4,5$. Features characteristic for the spectra of gapless bilayers can be immediately recognized in both systems. In the case of 5MLG/h-BN, we observe in addition the typical spectrum of MLG modified by h-BN layer (see inset with details of the band structure in the vicinity of the Brillouin-zon corner). It is worth noting that the band gap present in this spectrum is smaller than the one observed in TLG/h-BN system (compare insets of Fig.\ref{fig6} and \ref{fig7}).

\section{Conclusions }
Electrons in monolayer graphene act like massless spin-1/2 Dirac fermions. Backscattering is suppressed due to the pseudospin orthogonality of the forward and reverse scattering modes. The resulting Klein tunneling provides unit transmission for normally incident electrons at a pn junction, regardless of barrier height. In contrast, bilayer graphene electrons act like parabolic spin-1 systems with perfect reflection for normal incidence (anti-Klein tunneling). It has been revealed that Hamiltonian of Bernal stacking multilayer graphene systems can be block diagonalized into effective bilayer (BLG) and monolayer (MLG) Hamiltonians depending on parity of layer numbers \cite{b26}, \cite{b37}. Here, we have extended this procedure to decomposition of the Hamiltonian matrix describing multilayer graphene systems supported by h-BN layer, i.e. non-symmetric stacking structures. 

We found that the $N$-layer graphene Bernal stacking system deposited on h-BN layer can be described by $N/2$ isolated effective bilayer systems with one effective h-BN layer if N is even or by $(N-1)/2$ effective bilayers plus one MLG/h-BN effective bilayer if $N$ is odd. The decomposition procedure revealed non-trivial result connected with the effective on-site potentials $\varepsilon$ induced in graphene layers. These parameters act as compensation against the imposed asymmetry by h-BN support and show physical mechanism what happens in real systems. Moreover, taking into account the impact of h-BN on graphene properties seems to be closer to experimental situation than neglecting it. As examples, we compare the tunneling properties of effective BLG which can be recognized in different multilayer systems, both homo- and heterogeneous. 

Although slightly, the h-BN layer modifies the electronic properties of graphene layers present in the system. Thus, it can be regarded not only as insulating support ensuring pure graphene transport channels but equally can promote specific BLG's properties. This may pave the way for future applications for pseudo-spintronics with bilayer graphene \cite{b49}, \cite{b50}. The unique electronic structure of graphene systems can be used to create pseudo-spin analogous of giant magnetoresistance and other established spintronics effects \cite{b51}.\\

{\bf Acknowledgements}\\
 
This work is supported in part by University of Lodz.

\end{document}